\documentstyle[12pt]{article}
\newcommand{\bq}{\begin{equation}}
\newcommand{\eq}{\end{equation}}
\newcommand{\bqa}{\begin{eqnarray}}
\newcommand{\eqa}{\end{eqnarray}}
\newcommand{\ra}{\rightarrow}
\def\L{\Lambda}

\def\half{{1 \over 2}}
\def\s{\sigma}

\def\bC{{\bf C}}
\def\D{\Delta}
\def\a{\alpha}
\def\b{\beta}
\def\g{\gamma}

\def\l{\lambda}
\def\tp{2\pi i}
\def\ov{\over}
\def\pr{\prime}
\def\rt{\sqrt{2}}
\def\ra{\rightarrow}

\def\2pi{1\over 2\pi i}
\def\q{q-q^{-1}}
\def\.{\mathaccent''005F}
\def\^{\mathaccent''007E}
\def\~{\tilde}
\def\newline{\hfil\break}

\def\ra{\rightarrow}

\def\sq2{\sqrt{2}}
\def\sqk2{\sqrt{2(k+2}}
\def\sqk{\sqrt{k}}
\def\sqs{\sqrt{2\over k}}

\def\be{\begin{equation}}
\def\ee{\end{equation}}
\def\br{\begin{array}}
\def\er{\end{array}}
\def\bea{\begin{eqnarray}}
\def\eea{\end{eqnarray}}
\newcommand{\uq}{U_q (su(2)_k)}
\newcommand{\ou}{U_q (su(2)_1)}

\def\cF{{\cal{F}}}
\def\vp{\Omega}
\def\vP{\Phi}
\def\vt{\tilde{\Phi}}
\def\vo{\phi}
\def\cvo{\hat{\phi}}
\begin{document}
%%%%%%%%%%%%%%%%%%%%%%%%%

%% TITLE PAGE
\begin{titlepage}
\rightline{CRM-1875}
\rightline{hep-th/9305127}
\rightline{May 11, 1993}
\vbox{\vspace{15mm}}
\vspace{1.0truecm}
\begin{center}
{\LARGE \bf Matrix Elements of $U_q(su(2)_k)$ Vertex Operators via
Bosonization}\\[8mm]
{\large A.H. BOUGOURZI$^{1}$  and  ROBERT A. WESTON$^{2}$}\\
[3mm]{\it Centre de Recherche Math\'ematiques,
Universit\'e de Montr\'eal\\
C.P. 6128-A, Montr\'eal (Qu\'ebec) H3C 3J7, Canada.}\\[15mm]
\end{center}
\begin{abstract}
We construct bosonized vertex operators (VOs) and conjugate vertex operators
(CVOs) of
$U_q(su(2)_k)$ for arbitrary level $k$ and representation $j\leq k/2$.
Both are obtained
directly as two solutions of the defining condition
of vertex operators - namely that
they intertwine $U_q(su(2)_k)$ modules. We construct the screening charge
and present a formula for the n-point function.
Specializing to $j=1/2$ we construct all VOs and CVOs explicitly.
The existence of the CVO allows us to place the calculation of
the two-point function on the same footing as $k=1$; that is,
it is obtained without screening currents and involves only a single
integral from the CVO. This integral is evaluated and the resulting
function is shown to obey the q-KZ equation and to reduce simply to
both the expected $k=1$ and $q=1$ limits.

\end{abstract}
\footnotetext[1]{
Email: {\tt bougourz@ere.umontreal.ca}
}
\footnotetext[2]{
Email: {\tt westonr@ere.umontreal.ca}
}
\end{titlepage}

%% INTRODUCTION
\section{Introduction \label{secv1}}

In the last decade much effort has been expended in the study of two
dimensional exactly solvable models. The use of infinite-dimensional
conformal and chiral symmetries has proved eminently successful in the
 task of solving and classifying those models which are
critical\cite{bpz}. Following the
success of this program attention has shifted in recent years to massive
off-critical models. There are several  approaches
to studying these models including the Bethe Ansatz\cite{takfad}, the QISM
program of the Leningrad school\cite{qism} and the exact S-Matrix bootstrap
technique
\cite{zamsmatrix}.
The method that is most similar to that used in the
critical case is to make use of quantum or quantum-affine symmetries
\cite{Dri86}.
In analogy to CFT, the hope of this method is that it will enable one
to  describe the energy
levels, physical states,
correlation
functions, and in addition the S-matrix, of a lattice model or massive
quantum field theory
purely in terms of the representation theory of the quantum algebras.
In the recent works of the Kyoto school\cite{Daval92,collin,rsos,qaffine,
corner},
many of these goals
have been
realised for a class of quantum-spin chains, or related vertex models,
that are invariant under the action of $\uq$.

The Hamiltonian of the spin $1/2$ XXZ Heisenberg quantum
spin chain is \cite{affleck,takfad}
\be H_{XXZ}=-\half \sum_{i=-\infty}^{\infty} (\s_i^x \s_{i+1}^x
+\s_i^y \s_{i+1}^y + \Delta\s_i^z \s_{i+1}^z) \label{hamiltonian}.\eq
As a function of the anisotropy parameter $\Delta$, the phase
diagram of this model has three regions: for $\Delta<1$ there is a
massive antiferromagnetic phase; for $-1\leq \Delta \leq 1$ a massless
phase, and for $\Delta>1$ a massive ferromagnetic phase\cite{affleck,baxter}.
It is the massive antiferromagnetic
region that is considered in references \cite{Daval92,collin}. In
\ref{hamiltonian} the $\s_i$ are the Pauli matrices which act on the $i$th
component of the infinite tensor product
\be \cdots \otimes V \otimes V \otimes V \cdots, ~~~V=\bC^2. \label{infproduct}
\ee
When $\Delta=(q+q^{-1})/2$ the Hamiltonian is invariant under the
$\ou$ symmetry in the sense that
\be [\Delta^{\infty}(x),H]=0, \ee
where $\Delta^{\infty}(x)$ is the infinite coproduct
which defines formally how $x \in \uq$
acts on \ref{infproduct}.
Following \cite{FrRe92},
Davis et al.  introduced the following vertex operators that
intertwine $\uq$ modules\cite{Daval92}
\bq \vP^{j~j_2}_{~~j_1}:V(j_1) \ra V(j_2) \otimes V^j(z).
\eq
Here $V(j_i)$ is a spin $j_i$ $\ou$ module, $z$ is a spectral parameter
and $V^j(z)$ is
the $k=0$ `evaluation representation' of $\uq$ which is isomorphic to
$V^j\otimes {\bf C}[z,z^{-1}]$. Constructing the vertex operators
explicitly through
bosonization allowed Jimbo et al. to produce an integral formula for n-point
correlation functions of local fields\cite{collin}. In particular,
they calculated explicitly
the correlation function $<\s^z>$ (on alternate sites),
which triumphantly reproduced
the known formula for
the staggered polarization of the 6-vertex model
\cite{baxter} (the spin 1/2 chain and the 6-vertex model are equivalent
in that their transfers matrices commute\cite{baxter}).

For $k>1$ the above procedure can be repeated. The corresponding
physical systems are the spin $k/2$ XXZ chain\cite{Resh} or the spin $k/2$
generalisation of the 6-vertex model\cite{KulResh}. However the bosonization
procedure is more complex for $k>1$
\cite{bou93,Mat921,Mat922,Shi92,Bou92,BoGr92,Abaal92,Kim92}.
A trivial reason is that three
deformed bosonic fields are now necessary, as is the case for the
Wakimoto bosonization of $su(2)_k$\cite{Wak86}. For $k=1$\cite{Daval92},
the Frenkel-Jing
bosonization\cite{FrJi88}
was used which, like its undeformed ancestor the Frenkel-Kac
bosonization\cite{FrKa80},
requires only one field. A more serious difficulty is that in calculating
matrix elements of vertex operators one must balance the charge of the
bosonic out
vacuum (this charge is zero for $k=1$ as in CFT).
This can be done formally by inserting screening currents directly
into matrix elements of vertex operators\cite{Katal92,Mat922}.
However on its own this approach maximises the number of integrals
in the expression
for the n-point function.
The technique employed in CFT to overcome this, developed by Dotsenko, Fateev
and others\cite{DoFa84,Dot90,Mar89}, is to introduce conjugate vertex
operators.  Inserting  conjugate vertex operators, as well as vertex
operators, into the n-point function decreases the number of
screening currents required.

In this paper, we introduce the vertex operators and
conjugate vertex operators relevant to
$\uq$.
We are able to construct the bosonized forms of both of them.
They appear at the same time, as two possible solutions of the intertwining
requirement.
This construction is carried out for arbitrary $k$ and $j\leq k/2$.
After obtaining the  screening charges we present the formula
for n-point matrix elements. Specialising to $j=1/2$, we construct
the vertex operators and conjugate vertex operators explicitly.
The existence of the conjugate vertex operator allows us to place
the calculation of the two-point function on the same footing
as the $k=1$ case\cite{Daval92} - namely that the
function can be obtained in its simplest
form, which involves no screening currents and therefore no integrals.
We show that this two-point function is a solution
of the q-KZ equation.

The layout of the paper is as follows.
In Section 2, we define our bosonization of $\uq$ and the properties of
vertex operators. We present the intertwining condition and solve it
for arbitrary $k$ and $j$ to give the vertex operators and their conjugates.
In Section 3, we define the Fock space and the dual Fock space, and
discuss the role of BRST charges.
In Section 4, we introduce the screening charges, and present
the formula for n-point matrix elements.
In Section 5, we specialise to $j=1/2$, and construct the two vertex
operators and two conjugates vertex operators explicitly.
We calculate the two-point function, and show that it obeys
the q-KZ equation.
Finally we present
some conclusions and discuss the connection with quantum spin-chains.

\section{Bosonization of $U_q(su(2)_k)$\label{secbos}}
The associative $\uq$ algebra is generated by the operators
$\{ E^{\pm}_n~(n\in {\bf Z}),H_m ~(m \in {\bf Z} \neq 0),q^{\pm \rt H_0},
\g^{1/2}\}.$
In the Cartan-Weyl basis it reads \cite{Abaal92,Dri85,Jim85}
\be\br{rcl}
&&{[H_n,H_m]} = {[2n]\over 2n} {{\g^{nk}-\g^{-nk}} \ov {q-q^{-1}}}
\delta_{n+m,0},\qquad
n\neq 0,\\
&&{[q^{\pm\rt H_0}, H_m]}=  0,\\ && {[H_n,E^{\pm}_m]}=
\pm\sqrt{2}{\g^{\mp |n|k/2}[2n]\over 2n}
E^\pm_{n+m}, \qquad n\neq 0,\\
&&  q^{\rt H_0}  E^\pm_n q^{-\rt H_0} =q^{\pm 2} E^\pm_n,\\
&&{[E^+_n,E^-_m]} = {\g^{k(n-m)/2}\psi_{n+m}-\g^{k(m-n)/2}
\varphi_{n+m}\over q-q^{-1}},\\
&&E^\pm_{n+1}E^\pm_m-q^{\pm 2}E^\pm_mE^\pm_{n+1}=
q^{\pm 2}E^\pm
_nE^\pm_{m+1}-E^\pm_{m+1}E^\pm_n,
\label{cwb}
\er\ee
where $\g^{1/2}$ is in the centre of the algebra, and
as usual $[x]=(q^x-q^{-x})/(q-q^{-1})$. $\psi_n$ and $\varphi_n$ are
the modes of fields
$\psi(z)$ and $\varphi(z)$ defined by
\be\br{rcl}
\psi(z)&=&\sum\limits_{n\geq 0}\psi_nz^{-n}=q^{\sqrt{2}H_0}
 \exp\{\sqrt{2}(\q)\sum\limits_{n>0}H_nz^{-n}\},\\
\varphi(z)&=&\sum\limits_{n\leq 0}\varphi_nz^{-n}=q^{-\sqrt{2}H_0}
\exp\{-\sqrt{2}(\q)\sum\limits_{n<0}H_nz^{-n}\}.
\label{algebra}\er\ee
The above algebra is in fact only a subalgebra of
$\uq$ which includes in addition the grading operator and the Serre relations.
In what follows it will be clear from the context which algebra is being
considered.
The defining relations \ref{algebra} are the Drinfeld realization of
the usual $U_q(su(2)_k)$ which is given in the
Chevalley basis as:
\be\br{rcl}
&&t_it_j=t_jt_i,\\
&&t_i e_i t^{-1}_i=q^2e_i,\quad t_i e_j t^{-1}_i=q^{-2}e_j, ~~i\neq j,\\
&&t_i f_i t^{-1}_i=q^{-2}f_i,\quad t_i f_j t^{-1}_i=q^2f_j, ~~i\neq j,\\
&&[e_i, f_j]=\delta_{i,j}{t_i-t_i^{-1}\over q-q^{-1}},
\er\ee
where $i,j=0,1$.
Here the Chevalley generators $\{e_i,f_i,t_i\}$
are related to \ref{cwb} through the relations
\be\br{rcl}
t_0&=\gamma^k q^{-\rt H_0},\quad t_1&=q^{\rt H_0},\\
e_0&=E^-_1q^{-\rt H_0},\quad e_1&=E^+_0,\\
f_0&=q^{\rt H_0}E^+_{-1},\quad f_1&=E^-_0.
\er\ee
The above algebra is a Hopf algebra with the following comultiplication:
\be\br{rcl}
\Delta(e_i)&=&e_i\otimes 1+t_i\otimes e_i,\\
\Delta(f_i)&=&f_i\otimes t_i^{-1} +1\otimes f_i, \\
\Delta(t_i)&=&t_i\otimes t_i.
\er\ee
This comultiplication gives rise to the following comultiplication of the
$U_q(su(2)_k)$ algebra as realized in \ref{cwb}:
\be\br{rcl}
\!\!\!\!\!\!& &\Delta(E^+_n)=E^+_n\otimes\gamma^{kn}+
\gamma^{2kn}q^{\sq2 H_0}\otimes
E^+_n+ \sum_{i=0}^{n-1}\gamma^{k(n+3i)/2}\psi_{n-i}\otimes \gamma^{k(n-i)}
 E^+_i\: {\rm mod}   \: {N_-}\otimes {N_+^2},\\
%%%%%%%%%%%%%%
 \!\!\!\!\!\!& &\Delta(E^+_{-m})=E^+_{-m}\!\otimes\!\gamma^{-km}\!+\!
q^{-\sq2 H_0}\!\otimes\! E^+_{-m}+
\sum_{i=0}^{m-1}\gamma^{{k(m-i)\over 2}}\varphi_{-m+i}\otimes \gamma^{k(i-m)}
 E^+_{-i}
\: {\rm mod  }\:
 N_-\otimes N_+^2,\\
%%%%%%%%%%%%%%
\!\!\!\!\!\!& &\Delta(E^-_{-n})=E^-_{-n}\!\otimes\!\gamma^{-2kn}q^{-\sq2
H_0}\!+\!
  \gamma^{-kn}\!\otimes \!E^-_{-n}\!+\!\!\sum_{i=0}^{n-1}\!\gamma^{-k(n-i)}
  E^-_i\!\otimes\!\gamma^{{-k(n+3i)\over 2}}\varphi_{i-n}\: {\rm mod  }\:
 N_-^2\!\otimes\! N_+,\\
%%%%%%%%%%%%%%
\!\!\!\!\!\!& &\Delta(E^-_m)=\gamma^{km}\otimes E^-_m+E^-_m\otimes
q^{\sq2 H_0} +
\sum_{i=1}^{m-1}\gamma^{k(m-1)}E^-_m\otimes \gamma^{-k(m-i)/2}
\psi_{m-i}\:{\rm mod  }\: N_-^2\otimes N_+,\\
%%%%%%%%%%%%%
\!\!\!\!\!\!& &\Delta(H_m)=H_m\otimes\gamma^{km/2}+
\gamma^{3km/2}\otimes H_m\: {\rm mod  }\: N_-\otimes N_+,\quad\\
\!\!\!\!\!\!& &\Delta(H_{-m})=H_{-m}\otimes\gamma^{-3km/2}+\gamma^{-km/2}
\otimes  H_{-m}\: {\rm mod  }\: N_-\otimes N_+,\quad\\
\!\!\!\!\!\!& &\Delta(q^{\pm \sq2 H_0})=q^{\pm \sq2 H_0}\otimes
 q^{\pm \sq2 H_0},\quad\\
\!\!\!\!\!\!& &\Delta(\gamma^{\pm \half})=\gamma^{\pm\half}\otimes
\gamma^{\pm \half},
%$
\label{comult}\er\ee
where $m>0$, $n\geq 0$, and $N_\pm$ and $N_\pm^2$ are
left ${\bf Q}(q)[\gamma^\pm,  \psi_m, \varphi_{-n}; \: m, n\in {\bf
Z_{\geq 0}}]$ modules
generated
 by $\{E^\pm_m; \:m\in {\bf Z}\}$
and $\{E^\pm_m E^\pm_n; \:m, n\in {\bf Z}\}$  respectively
\cite{Jimal92,ChPr91}.
This comultiplication will be useful in deriving the intertwining
properties of the
vertex operators.
For the purpose of bosonization, it is convenient to rewrite this
quantum algebra in terms of OPEs (the quantum current algebra).
The $U_q(su(2)_k)$ QCA
then reads  \cite{Abaal92,Mat921,Mat922,Ber89}
\bea
\psi(z).\varphi(w)&=&
{(z-wq^{2+k})(z-wq^{-2-k})\over (z-wq^{2-k})(z-wq^{-2+k})}
\varphi(w).\psi(z),
\label{ope4}\\
\psi(z).E^{\pm}(w)&=&
q^{\pm 2}{(z-wq^{\mp(2+k/2)})\over z-wq^{\pm (2-k/2)}}
E^\pm(w).\psi(z),
\label{ope5}\\
\varphi(z).E^{\pm}(w)&=&
q^{\pm 2}{(z-wq^{\mp(2-k/2)})\over z-wq^{\pm (2+k/2)}}
E^\pm(w).\varphi(z),
\label{ope6}\\
E^+(z).E^-(w)&\sim& {1
\over w(\q)}\left\{{\psi(wq^{k/2})\over z-wq^k}-
{\varphi(wq^{-k/2})\over z-wq^{-k}}\right\},\quad |z|>|wq^{\pm k}|,
\label{ope7}\\
E^{\pm}(z).E^{\pm}(w)&=&{(z q^{\pm 2}-w)\over z-w q^{\pm 2}}
E^{\pm}(w). E^{\pm}(z).
\label{ope8}
\eea
\subsection{Bosonization of the currents\label{subwaki}}

In order to bosonize the ``classical'' $su(2)_k$ current algebra for
arbitrary $k$,
it is necessary to introduce
three sets of bosonic oscillators \cite{bou93}.
This is known as the Wakimoto bosonization \cite{Wak86}.
For $k=1$ however, the Frenkel-Kac
bosonization is also available - which uses only one set of
oscillators\cite{FrKa80}.
For the quantum $\uq$, the situation is analogous. For $k=1$ there is
the Frenkel-Jing bosonization which requires only one set of deformed
oscillators\cite{FrJi88}, and for general $k$ there
are (at least five) q-deformations
of the Wakimoto bosonization available
\cite{bou93,Mat921,Mat922,Shi92,Abaal92,BoGr92,Kim92}.
(See \cite{bou93} for a detailed
discussion of the equivalence of these different
bosonizations.)  We shall use the fifth bosonization
introduced in \cite{bou93}.

The three deformed Heisenberg algebras required are
\be\br{rcl}
{[a^j_n,a^\ell_m]}&=&(-1)^{j-1}nI_j(n)\delta^{j,\ell}
\delta_{n+m,0},\\
 {[a^j,a^\ell_0]}&=&(-1)^{j-1}i\delta^{j,\ell},
\qquad\qquad\qquad  j,\ell=1,2,3,
\er\ee
where
\be\br{rcl}
I_1(n)&=&{[2n][nk]\over 2kn^2},\\
I_2(n)&=&{[nk][n(2+k)]\over n^2k(2+k)}q^{nk},\\
I_3(n)&=&{[2n]^2\over 4n^2}.
\er\ee
Inspired by  the notation of Kato et al. \cite{Katal92}, we choose to express
all of our fields in terms of three generic fields
$\vp^j(L,M,N|\s,\a,\b|\pm|z)~~\{j=1,2,3\} $ defined as follows:
\be\br{lll}
\vp^j(L,M,N|\s,\a,\b|\pm|z)&=a^j-ia_0^j\ln{(\pm zq^\s})\\
&+i {{LM}\over N}\sum\limits_{n>0} {{[Nn]}\over {[Ln][Mn]}}
a_n^j q^{\a n} z^{-n}
+i {{LM}\over N}\sum\limits_{n<0} {{[Nn]}\over {[Ln][Mn]}}
a_n^j q^{\b n} z^{-n}.
\er\ee
where $L$, $M$, $N$, $\sigma$, $\alpha$ and $\beta$ are parameters
associated with the q-deformation. These deformed
fields are
normalized so that they coincide with usual free bosonic field in the limit
$q\ra 1$.
All OPEs are then given in terms of expression \ref{OPE} in the appendix.

Let us define the deformed bosonic fields
\be\br{rcl}
\chi^{1,\pm}(z)&=&\vp^1(k,1,1|0,\mp k/2,\pm k/2|+|z),\\
\chi^{2}(z)&=&\vp^2(k,1,1|0,-k/2,k/2|+|z),\\
\chi^{3}(z)&=&\vp^3(2,1,1|0,k+2,k+2|+|z).
\er\ee
In terms of these fields the bosonization of the currents is
\be\br{rcl}
\!\!\!\!\psi(z)&= &\exp\left\{
i\sqs\left(\chi^{1,+}(zq^{k/2})-\chi^{1,-}(zq^{-k/2})\right)
\right\}\\
&=&q^{\sqrt{2k}\chi^1_0}
\exp\left\{\sqrt{2k}(\q)\sum_{n>0}\chi^1_nz^{-n}\right\},\\
\!\!\!\!\varphi(z)&= &\exp\left\{
i\sqs\left(\chi^{1,+}(zq^{-k/2})-\chi^{1,-}(zq^{k/2})\right)
\right\}\\
&=&q^{-\sqrt{2k}\chi^1_0}
\exp\left\{-\sqrt{2k}(\q)\sum_{n<0}\chi^1_nz^{-n}\right\},\\
\!\!\!\!E^{+}(z)&=&{\exp\{i\sqs\chi^{1,+}(z)+i\sqrt{{2+k\over k}}
\chi^2(z)\}\over z(\q)}
\left(\exp\{-i\chi^3(zq^{-1})\}
-\exp\{- i\chi^3(zq)\}\right)\\
&\equiv& {1 \over {z(q-q^{-1})}} (E_1^{+}(z)-E_2^{+}(z)),\\
\!\!\!\!E^{-}(z)&=&{\exp\{- i\sqs\chi^{1,-}(z)
\}\over z(\q)}\left(\exp\{-i\sqrt{{2+k\over k}}
\chi^2(zq^k)+i\chi^3(zq^{1+k})\}\right.\\
&&\left.-\exp\{-i\sqrt{{2+k\over k}}
\chi^2(zq^{-k})+ i\chi^3(zq^{-1-k})\}\right)\nonumber\\
&\equiv& {1 \over {z(q-q^{-1})}} (E_1^{-}(z)-E_2^{-}(z)).\\
\label{fif}
\er\ee
In these, and in all other expressions in this paper,
operators and products of operators defined at the same point $z$
are understood to be normal ordered with respect to the
Heisenberg generators.

\subsection{The vertex operators and their conjugates}
The vertex operators relevant to this discussion are
the intertwiners  of Section 6 of reference\cite{Daval92}.
They are defined as maps between $\uq$ modules  in
the following way
\bq \vP^{j~j_2}_{~~j_1}(z):V(j_1) \ra V(j_2) \otimes V^j(z).
\label{origvo}\eq
Here $V(j)\equiv V(\Lambda_j)$ are left highest weight $\uq$
modules, with $\{\Lambda_j=(k-2j)\l_0+2j\l_1,~ j=0,\dots,k/2\}$
and $\{\l_0,\l_1\}$ denoting the sets of $\uq$ dominant highest
weights and fundamental weights respectively. $V^j(z)$ is the
$k=0$ `evaluation representation' of $\uq$.
It is isomorphic to $V^j\otimes {\bf C}[z,z^{-1}]$, where $V^j$ is the
$2j+1$ dimensional
representation with the basis $\{v_m^j,~ -j\leq m\leq j\}$.
$V^j(z)$ is equipped with the following $U_q(su(2)_k)$ module
structure \cite{Katal92}:
\be\br{rcl}
\gamma^{\pm 1/2}v^j_m&=&v^j_m,\\
q^{\rt H_0} v_m^j&=&q^{-2m}v^j_m,\\
E^+_nv^j_m&=&z^nq^{2n(l-m)}[j-m+1]v^j_{m-1},\\
E^-_nv^j_m&=&z^{n}q^{-2nm}[j+m+1]v^j_{m+1},\\
H_nv^j_m&=&{z^n\over{\rt n}}\{[2nj]-q^{n(j-m+1)}(q^n+q^{-n})[n(j+m)]\}v^j_m,
\er\ee
where is it understood that $v^j_m$ is identically zero if $m> j$ or
$m< -j$.
We also define vertex operators
$\vt^{j~j_2}_{~~j_1}(z)$ as
\be \vP^{j~j_2}_{~~j_1}(z)=z^{(\D_{j_2}-\D_{j_1})}
\vt^{j~j_2}_{~~j_1}(z),\ee
where $\D_j= j(j+1)/(k+2)$. By definition these vertex operators
obey the intertwining condition\cite{FrRe92,Daval92}
\bq \vt^{j~j_2}_{~~j_1}(z) \circ x = \Delta(x)\circ
\vt^{j~j_2}_{~~j_1}(z) ~~~~~\forall~ x\in \uq .
\label{int} \ee
Here $x$ denotes a generator in the Drinfeld realization,
which is appropriate for
explicitly constructing the vertex operators in terms of free bosons.
It is also convenient to define components of these vertex operators through
\be \vt^{j~j_2}_{~~j_1}(z) = g^{j~j_2}_{~~j_1}(z)
\sum\limits_{m=-j}^{j}  \vo^j_m(z)\otimes v^{j}_m ,\ee
where the normalisation function $g^{j~j_2}_{~~j_1}(z)$ is given by
\be g^{j~j_2}_{~~j_1}(z)=(-zq^{k+2})^{(\D_j+\D_{j_1}-\D_{j_2})}.\ee

Using the above relation \ref{int}, the
comultiplication \ref{comult}, and the fact that
$N_+V^j_{-j}=N_-V^j_j=0$, $N_{\pm}V^j_m\in
F[z,z^{-1}]
V^j_{m\mp 1}$, we arrive at the following commutation relations:
\pagebreak
\bea
{[E^{+}(w),\vo^j_j(z)]}&=&0,\\
{[H_n, \vo^j_j(z)]}&=&j\rt \left\{\delta_{n,0}+
q^{(n(k+2)+|n|k/2)} { {[2jn]}\ov {2jn} }z^n \right\}\vo^j_j(z),\\
\vo^j_m(z)&=&{1\over {[j-m]!}}[\dots[\vo^j_j(z),E^{-}_0]\dots,
E^-_0]_{q^{2(m+1)}}.
\label{phijm}\eea
In \ref{phijm} there are $(j-m)$ quantum commutators,
where the quantum commutator $[A,B]_{q^{x}}$ is defined by
\be\br{rcl}
[A,B]_{q^x}=AB-q^xBA.
\er\ee

Contrary to the case  of the Frenkel-Jing realization
(i.e. $k=1$)\cite{collin},
the system of equations \ref{int} has two independent solutions
$\vo^j_j(z)$ and $\cvo^j_j(z)$  given in
terms of the above bosonization by
\be\br{rcl}
\vo^j_j(z)&=&\exp\{j\sqrt{2\over k}i\xi^1(z)+{2j\over
\sqrt{k(k+2)}}i\xi^2(z)\},\\
\cvo^j_j(z)&=&X^j_j(z)\vo^j_j(z),\\
X^j_j(z)&=&\exp\{{-k(k+1-2j)\over \sqrt{k(k+2)}}i\hat\xi^2(z)+
{(k-2j)}i\hat\xi^3(z)\},
\er\ee
where
\be\br{rlc}
&\xi^1(z)=\vp^1(2,k,2j|k+2,-2-k/2,-2-3k/2|-|z),\\
&\xi^2(z)=\vp^2(k+2,k,2j|k+2,-2-3k/2,-2-k/2|-|z),\\
&{\hat\xi}^2(z)=\vp^2(k+2,1,k+1-2j|k+2,-2-3k/2,-2-k/2|-|z),\\
&{\hat\xi}^3(z)=\vp^3(2,1,k-2j|k+2,0,0|-|z).
\er\ee
$\vo^j_m(z)$ and $\cvo^j_m(z)$ are derived from
$\vo^j_j(z)$ and $\cvo^j_j(z)$ through \ref{phijm}. From now on we shall refer
to $\vo^j_m(z)$ and $\cvo^j_m(z)$ as the vertex operators (VOs) and
conjugate vertex operators (CVOs) respectively.

\section{ Fock modules and Fock spaces \label{secfoc}}

We  define the Fock module $F(n_1,n_2,n_3)$ and its dual
$F^{\dag}(n_1,n_2,n_3):F(n_1,n_2,n_3)\ra {\bf C}$ as follows:
\be\br{rcl}
F(n_1,n_2,n_3)&=&F_-|n_1,n_2,n_3>,\\
F_+|n_1,n_2,n_3>&=&0,
\er\ee
where $F_\mp$ is a free ${\bf Q}(q)$ module generated by
$\{a_{\mp n}^1,a_{\mp n}^2,a_{\mp n}^3,~ n>0\}$, and the states
$|n_1,n_2,n_3>$, labelled by integers $n_1$,
$n_2$ and $n_3$, are
defined by
\be\br{rcl}
|n_1,n_2,n_3>=\exp\{ {n_1  \over \sqrt{2k}}ia^1+{n_2 \over
\sqrt{k(k+2)}}ia^2+n_3 ia^3\}|0>,
\er\ee
where $|0>=|0,0,0>$ is the `in' vacuum and it is annihilated by
$\{a^1_n,a^2_n,a^3_n;~ n\geq 0\}$. As in the classical ($q=1$) case
\cite{Dot90},
there
will be an asymmetry between the `in' and the `out' vacua due to the existence
of a background charge. Guided as always by this example,
we define the out vacuum (and our normalisation) by
\be <0,k(k+1),-k|0>=1, \label{norm}\ee
where \be
<n_1,n_2,n_3|=<0|\exp\{ - {n_1  \over \sqrt{2k}}ia^1-{n_2 \over
\sqrt{k(k+2)}}ia^2-n_3 ia^3\}. \ee
The dual Fock module $F^{\dag}(n_1,n_2,n_3)\simeq F(n_1,n_2-k(k+1),n_3+k)$
will then be given formally
by
\be F^{\dag}(n_1,n_2,n_3)=<n_1,n_2+k(k+1),n_3-k|F_{+}.\ee

The currents $E^\pm(z)$,  VOs and CVOs define the following mappings on the
Fock modules:
\bea
\!\!\!E^{\pm}(z):& F(n_1,n_2,n_3)\ra F(n_1\pm 2,n_2\pm(k+2),n_3\mp 1),
\label{eact}\\
\!\!\!\vo^j_m(z):& F(n_1,n_2,n_3)\rightarrow F(n_1+2m,n_2+m(k+2)-jk,n_3+j-m),
\label{act2}\\
\!\!\!\cvo^j_m(z):& F(n_1,n_2,n_3)\rightarrow F(n_1+2m,n_2+m(k+2)-k(k+1-j),
n_3+k-j-m)
\label{action}\eea
One can show that the action of the currents is well defined (single valued)
on the Fock modules $F(n_1,n_2,n_3)$ and $F^{\dag}(n_1,n_2,n_3)$ provided
that the following
condition is satisfied:
\be\br{rcl}
n_1-n_2\in k{\bf Z}.
\er\ee
{}From \ref{eact} it is clear that the representations of the currents are the
complete Fock spaces  \bq \cF (n_1,n_2,n_3)=\bigoplus_{r\in{\bf Z}}
F(n_1+2r,n_2+r(k+2),n_3-r). \label{FS}\eq
Let us now briefly discuss the  embedding of  the left $U_q(su(2)_k)$ highest
weight modules $V^j\equiv
V(\Lambda_j)$ in the above Fock space $\cF(n_1,n_2,n_3)$.
The same analysis will apply to
the embedding of the right modules $V^{\dag}(\Lambda_j)$ in the Fock space
$\cF^{\dag}(n_1,n_2,n_3)$.
The $U_q(su(2)_k)$ highest weight states  $|j>\equiv|\L_j>$ must satisfy the
conditions:
\be\br{rcl}
e_i|\Lambda_j>=0,\quad i=0,1.
\er\ee
It can be easily checked that the state $|j>$ can be identified with the
highest weight state $|2j,2j,0>$ of  the Fock module $F(2j,2j,0)$.
As noted in \cite{Mat922}, and as is well known in the
classical case, one has to study the BRST  cohomology structure of
the Fock space $\cF(2j,2j,0)$
in order to single out the irreducible highest weight $V(j)$, with
$0\leq j\leq k/2$,
and show that
the vertex operators are acting in the latter, which then ensures the
nonvanishing of their matrix elements.
This analysis has been partially carried out
in Ref \cite{Mat922}.
Here we will conjecture that the classical result
(described fully in \cite{Fel??})
still holds in the quantum case - namely that we have the isomorphism:
\be\br{rcl}
V^j\simeq \delta_{s,0}KerQ^{2j+1}_s/ImQ^{k-2j+1}_{s+1},~~ 0\leq j\leq k/2,
{}~~s\in{\bf Z}, \label{iso}
\er\ee
where $Q^{2j+1}_s$ and  $Q^{k-2j+1}_s$  are  BRST  charges
acting in a graded complex of the quantum analogue of the Fock spaces
introduced in \cite{Fel??}.
Let us just mention here that contrary to the Fock spaces of \cite{Fel??},
the cohomology structure of the Fock spaces $\cF(n_1,n_2,n_3)$
given by \ref{FS} is
much more involved and requires the introduction of at least one extra
screening charge \cite{Mat922}.
The BRST charges of \ref{iso} are
constructed as integer powers of the screening charge $Q$, which is itself
constructed
in terms of a  screening current that  will be introduced below.  We also
conjecture that as in the classical case the VOs and
CVOs are constructed as BRST invariant combinations of free fields (see
\cite{Fel??}).

\section{The Screening Charge and Screening Current}

As is well know in CFT, a screening charge $Q$ is a dimensionless
operator that
commutes with
the underlying symmetry algebra\cite{bpz,DoFa84,Dot90}. Using the powerful
technics of OPEs it can be
systematically constructed as a closed contour integral of a dimension
1 operator,
called
a screening current $S(z)$, which commutes  with
the currents generating the algebra up to total derivatives.
In the case of $U_q(su(2)_k)$ one expects that  $S(z)$ will
commute with
the currents $E^\pm(z)$, $\varphi(z)$ and $\psi(z)$ up to total {\it quantum}
derivatives.
A quantum derivative of a function $f(z)$ is defined by:
\be\br{rcl}
_k{\cal D}_zf(z)={f(zq^k)-f(zq^{-k})\over z(q-q^{-1})}.
\er\ee
In
fact one can easily show that the screening current
\be\br{rcl}
S(z)~= ~{_1{\cal D}}_z(\exp\{-i{\eta}^3(z)\})
\exp\{i\sqrt{k\over k+2}{\eta}^2(z)\}
\er\ee
with
\be\br{rcl}
\eta^2(z)=\Omega^2(k+2,1,1|-k-2,-k/2,k/2|+|z),\\
\eta^3(z)=\Omega^3(2,1,1|-k-2,k+2,k+2|+|z),
\er\ee
satisfies the following commutation relations:
\bea
{[\varphi(z),S(w)]}&=&0,\\
{[\psi(z),S(w)]}&=&0,\\
{[E^+(z),S(w)]}&=&0,\\
{[E^-(z),S(w)]}&=&-_{k+2}{\cal D}_w({h(w)\over z-w}),
\label{ecoms}\eea
where $h(w)$ is  regular except at $w=0$.
Because of relation \ref{ecoms} the screening charge $Q$ is constructed
as a Jackson
integral of the
 screening current, that is,
\be\br{rcl}
Q=\int_0^{s\infty}d_pz S(z) \label{Q},
\er\ee
with $p=q^{2(k+2)}$. Here the Jackson integral of a function $f(z)$ is
defined by \cite{FrRe92}:
\be
\int_0^{s\infty}d_pz f(z)=s(1-p)\sum\limits_{n\in {\bf Z}}f(sp^n)p^n. \ee

The action of the screening charges \ref{Q} on $F(n_1,n_2,n_3)$ is given
by
\be Q:F(n_1,n_2,n_3) \ra F(n_1,n_2+k,n_3-1)\label{qact}.\ee
The two BRST charges introduced in the previous section are given in terms of
this screening charge by $Q^x_s=Q^x$,
where $x=2j+1$ or $x=k-2j+1$ depending on whether $Q^x_s$ is acting in
$\{\cF(j+s(k+2)),~ s\in {\bf Z}\}$ or
$\{\cF(-j-1+s(k+2)),~ s\in {\bf Z}\}$ respectively. These particular values of
$x$ ensure that the BRST charges are single valued when acting on these
Fock spaces.

The screening charges are necessary in order to calculate matrix elements.
They must be inserted in the correct number in order to balance the charge
of the vacuum.
The n-point matrix element
$<{\hat\vP}^{j_n}(z_n)\vP^{j_{n-1}}(z_{n-1})\cdots\vP^{j_1}(z_1)>$  of the
original vertex operators of \ref{origvo} is given by
\be\br{lll}&<{\hat\vP}^{j_n}(z_n)\vP^{j_{n-1}}(z_{n-1})
\cdots\vP^{j_1}(z_1)>=\\
&\!\!\!\!\!\!\!\!\sum\limits_{\{m_1,\cdots,m_n\}} \left(\prod\limits_{i=1}^{n}
z_i^{\D_{J_i}-\D_{J_{i-1}}} g^{j_i~J_i}_{~~J_{i-1}}(z_i)\right)
<0|Q^L\cvo^{j_n}_{m_n}(z_n) \vo^{j_{n-1}}_{m_{n-1}}(z_{n-1}) \cdots
\vo^{j_1}_{m_1}(z_1)
|0> v_{m_n}^{j_n}\otimes \cdots \otimes v_{m_1}^{j_1},\label{npoint}
\er\eq
where the $J_i$ indices are specified by  $\{J_0=0,~J_i=\sum_{\ell=1}^{i}
j_{\ell},~J_n=J_{n-1}-j_n\}$. $J_i$ and $J_{i-1}$ label the highest weight
modules that are intertwined by a VO or CVO with the argument $z_i$. They
are fixed by the action of \ref{act2} and \ref{action}, and
are suppressed from the
left hand side above.
The requirement that one must balance the vacuum
charge is equivalent to the condition that after  acting with
the VOs, CVOs and Qs one arrives at a product \linebreak
$<n_1,n_2+k(k+1),n_3-k|n_1,n_2,n_3>={\bf C}$ for some $n_i$.
This dictates that firstly the
sum extends over only
those $-j_i \leq m_i \leq j_i$ such that $\sum_{i=1,n} m_i=0$,
and secondly that
the number of screening currents is $L=-j_n+\sum_{i=1,n-1} j_i$.
Additional constraints are provided by demanding that the correlation function
be independent of the position at which the CVO is inserted. If it is placed
in the $\ell{\rm{th}}$ position instead of the first, then $L_{\ell}=-j_{\ell}
+\sum_{i\neq\ell}j_i$ screening currents are required. The requirement
that $(L_i\geq0;~i=1,\dots,N)$ fixes the ``fusion rules" for
the vertex operators \cite{Alval89}.

As an example of the above constraints, consider the two-point function.
Depending on the position of the CVO the numbers of screening currents
required are $L_1=-j_1+j_2$ and $L_2=-j_2+j_1$. The conditions  $L_i\geq 0$
imply that
$j_1=j_2$ and $L_1=L_2=0$. Thus the only non-vanishing terms in \ref{npoint}
are of the form
\be <\cvo^j_m(z)\vo^j_{-m}(w)>. \ee
%%%%%%%%%%%%%%%%%%%%%%%%%%%%%%%%%%%%%%%%%%%%%%%%%%%%%%%%%%%%
%%%SECTION
\section{j=1/2 Vertex Operators and Matrix Elements}

The explicit
solutions to
the intertwining conditions \ref{int} for $j=1/2$  are
\bea \vo_{\half}^{\half}(z) &=& \exp\left({1 \over {\sqrt{2k}}}
i\xi^1(z)+{1 \over {\sqrt{k(k+2)}}} i\xi^2(z)\right), \\
\vo_{-\half}^{\half}(z)&=&\oint\limits_{|w|<|zq^{1+k}|}
{dw \over \tp} {{:~\vo_{\half}^{\half}(z)E_2^{-}(w)~:}\ov{q(w-zq^{1+k})}}
-q \oint\limits_{|w|>|zq^{3+k}|} {dw \over \tp}
{{:~\vo_{\half}^{\half}(z)E_1^{-}(w)~:}\ov{(w-zq^{3+k})}}\label{vominus},\\
\cvo_{\half}^{\half}(z) &=&X(z)\vo_{\half}^{\half}(z),~~~~~~~\hbox{where}\\
X(z)&=&\exp\left(-{k^2 \ov {\sqrt{k(k+2)}}} i\xi^2(z) +(k-1)i\xi^3(z)\right),\\
\cvo_{-\half}^{\half}(z)&=&-(q-q^{-1})\oint\limits_{|zq^{k+3}|<|
w|<|zq^{1+k}|}
dw w{ {:~\vo^{\half}_{\half}(z)E^{-}(w)~: } \ov
{(w-zq^{k+1})(w-zq^{k+3})} }.
\label{cvominus}\eea

It is interesting to ask the question: why can't one simply integrate
\ref{vominus} and \ref{cvominus} as one does with the
analogous expressions for $su(2)_k$ vertex
operators\cite{Dot90}?
The answer is that in this later case ($q=1$) each of \ref{vominus} and
\ref{cvominus} becomes written as a difference of two integrals, with the
same integrands but different contours. The first contour includes only the
pole at $w=0$, whereas the second one includes both
the poles at $w=0$ and $w=z$. This means that each of \ref{vominus}
and \ref{cvominus} can be expressed in the classical case as a single
integral but with the contour now winding around the pole
$w=z$ only.
Or equivalently the
contributions of the two  integrals from the pole at $w=0$
cancel. This is no longer true in the quantum case because the two
integrands differ. Furthermore one cannot simply evaluate the residue
of the pole at  $w=0$ because its order is not fixed until we act on our
Fock space with the vertex operator. So these integrals
must be left until we come to evaluate a specific correlation function.

The two-point matrix element corresponding to the left hand side of
\ref{npoint} is
\be\br{lll} &<{\hat\vP}^{\half}(z)\vP^{\half}(w)>= \\
&(w/z)^{\half} g^{\half~0}_{~~\half}(z)
g^{\half~\half}_{~~0}(w)
(<0|\cvo^{\half}_{\half}(z)\vo^{\half}_{-\half}(w)|0>
v_{+}\otimes v_{-}+\\
&<0|\cvo^{\half}_{-\half}(z)\vo^{\half}_{\half}(w)|0>
v_{-}\otimes v_{+}),\er\ee
where $v_{\pm}$ correspond to $v^{\half}_{\pm \half}$.
Carrying out the integrals we evaluate this expression as
\be f({w / z})(v_{-}\otimes v_{+} - q v_{+}\otimes v_{-}),~~\hbox{where}~~\\
f(z)=z^{3\ov {4(k+2)}}\prod\limits_{n=0}^{\infty} { {(p^{n+1} z)_{\infty}
(p^{n+1} q^4 z)_{\infty}} \ov {(p^{n+1}q^{-2} z)_{\infty}
(p^{n+1} q^6 z)_{\infty}} } \label{fzw},\ee
and $p=q^{2(k+2)}$.
Here we have used the conventional notation
\be (a)_{\infty}=\prod\limits_{n=0}^{\infty}(1-aq^{4n}).\ee
It can be easily shown that the function $f(z)$ is a solution
of the q-KZ equation\cite{FrRe92}
\be f(pz)=q^{3\ov 2} { {(pq^{-2}z)_{\infty} (pq^6z)_{\infty}}
\ov {(pz)_{\infty} (pq^4z)_{\infty}} } f(z).\ee
Moreover when we set $k=1$, $f(z)$ becomes
\be f(z)=z^{1\ov 4}{ {(q^6z)_{\infty}} \ov {(q^4z)_{\infty}} }.\ee
This expression is that found by Davies et al.\cite{Daval92}
using the Frenkel-Jing bosonization
for $k=1$.
The function $f(z)$ can be expressed in the appealing exponential
form
\be f(z)=z^{3\ov {4(k+2)}}\exp \sum\limits_{n>0}
{ z^{n} \ov n}{ {[n][3n]}\ov{[2n][n(k+2)]} }
q^{n(k+2)}. \ee
The limits $k=1$ and $q \ra 1$ can both be read off simply from this form.
Setting $k=1$ gives the expression above. Taking the `classical'
limit $q \ra 1$ yields the expression
\be {{z^{\D_{\half}}} \ov { (1-z)^{2\D_{\half}} } },\ee
where $\D_{\half}=3/(4(k+2))$ is the
conformal weight expected for the $su(2)_k$ CFT
\cite{KnZa84}.

\section{Conclusions}

To summarise: In this paper we have constructed vertex operators and conjugate
vertex operators at the same level, as solutions of the intertwining
conditions \ref{int}. We have constructed the screening charge, and
given an expression for a general n-point matrix element function for
arbitrary level $k$. By demanding that the vacuum charge be balanced,
and that the n-point matrix elements should be invariant under changing the
position of the CVO we have obtained the fusion rules. Specialising to
$j=1/2$ we have constructed the VOs and CVOs explicitly. The existence
of the CVOs has allowed us place the calculation of the two-point function on
the same footing as the $k=1$ case \cite{Daval92,collin} - namely that the
two-point
function is obtained without screening currents and involves only a single
classical integral from the CVO. This integral has been carried out to produce
a
function \ref{fzw} that obeys the q-KZ equation, and reduces simply
to both the $k=1$ and $q=1$ limits.

What of the spin $k/2$ quantum spin chain, or related vertex model?
In order to calculate thermodynamic quantities such as the polarisation
or susceptibility of relevance to these models it is necessary to calculate
correlation functions of local fields. These fields can be
considered as acting  on the $V^j(z)$ representation; and n-point correlation
functions of local fields can be expressed in terms of 2n-point matrix
elements of the VOs. For $k=1$, a formula for general n-point
correlation functions was obtained in this manner by Jimbo et al.\cite{collin}.
The results described above allow for a rather natural generalisation of
the $k=1$ case. Details will be published elsewhere.

\section*{Acknowledgements}
We wish to thank Amine El Gradechi for many illuminating discussions.
We are also grateful to  Luc Vinet for providing us with his notes
on q-hypergeometric functions. RAW thanks CRM for
providing him with a research fellowship.

\newpage
%% THE APPENDIX
\appendix
\section{Appendix}
\subsection*{Correlation functions and OPEs}
In calculating the OPE of two vertex operators one uses
the usual expression \be  :~e^A~::~e^B~:=:~e^Ae^B~:e^{<AB>} ~~~.\ee
For strings of operators this generalises to `Wick's theorem for
vertex operators'. That is
\be \prod\limits_{i=1}^{N} :~e^{A_i}~:=:~\prod\limits_{i=1}^{N}~
{}~e^{A_i}~:
\exp(\sum\limits_{i<j}<A_iB_j>)~~~.\ee
Thus to calculate any of the OPEs and matrix elements used in this
paper it is sufficient to know the two-point function of our generic
field $\vp^{j}$.
This is given by the following expression
\be\br{rcl}& \exp\left(<\vp^j(L,M,N|\s,\a,\b|\ell|z)\vp^{j^{\pr}}
(L^{\pr},M^{\pr},N^{\pr}|\s^{\pr},\a^{\pr},\b^{\pr}
|\ell^{\pr}|z^{\pr})>\right) = \\
& (\ell zq^{\sigma})^{(-1)^j}\exp\left((-1)^{j-1}{{LML^{\pr}M^{\pr}} \ov
{NN^{\pr}} } \sum\limits_{n > 0}
{ {[Nn][N^{\pr}n]nI_j(n)} \ov {[Ln][Mn][L^{\pr}n][M^{\pr}n]} }
q^{(\a-\b^{\pr})n} ({z^{\pr} \ov z})^n \right).\label{OPE}\er\ee
Using the two identities \be \sum\limits_{n=0}^{\infty}x^n =
{ 1 \ov {(1-x)}}~~\hbox{for}~~|x|<1, \ee
and
\be \sum\limits_{n=1}^{\infty} { x^n \ov n}=-\ln(1-x) ~~\hbox{for}~~|x|<1, \ee
in various combinations then allows the sum to be carried out.

\end{document}